\documentclass[10.5pt,twocolumn,showpacs,showkeys,preprintnumbers,amsmath,amssymb,prb,superscriptaddress]{revtex4}
\usepackage[utf8]{inputenc}
\usepackage{eqnarray,amsmath}
\usepackage{amsfonts}
\usepackage{epsfig}
\usepackage{bm}
\usepackage{setspace}
\usepackage{booktabs} 
\usepackage{graphicx}
\usepackage[colorlinks=true, citecolor=red, linkcolor=blue, urlcolor=blue]{hyperref}
\usepackage{booktabs}
\usepackage{titlesec}
\titlespacing*{\section}
{0pt}{1.5ex plus 1ex minus .2ex}{3.3ex plus .2ex}
\titlespacing*{\subsection}
{0pt}{1.5ex plus 1ex minus .2ex}{3.3ex plus .2ex}       
\begin{document}
\title{Structures and infrared spectroscopy of Au$_{10}$ cluster at different temperatures}
\author{Francisco Eduardo Rojas-Gonz\'alez}
\affiliation{Departamento de F\'isica, Edificio 3F, Universidad de Sonora, Blvd. Luis Encinas y Rosales S/N, 83000 Hermosillo, Sonora, M\'exico.}
\author{Jorge Padilla-Alvarez}
\affiliation{Departamento de Ciencias Tecnol\'ogicas del Centro Universitario de la Ci\'enega de la Universidad de Guadalajara. Av. Universidad, N\'um.1115, Col. Lindavista, Ocotl\'an, Jalisco, M\'exico.}
\author{C\'esar Castillo-Quevedo}
\affiliation{Departamento de Fundamentos del Conocimiento, Centro Universitario del Norte, Universidad de Guadalajara, Carretera Federal No. 23, km 191, C.P. 46200, Colotl\'an, Jalisco, M\'exico.}
\author{Rajagopal Dashinamoorthy Eithiraj}
\affiliation{Department of Physics, School of Advanced Sciences, Vellore Institute of Technology (VIT), Chennai 600127, Tamil Nadu, India }
\author{Jose Luis Cabellos}\email[email:]{sollebac@gmail.com, jose.cabellos@uptapachula.edu.mx}
\affiliation{Coordinaci\'on de investigaci\'on y desarrollo tecnol\'ogico, Universidad Polit\'ecnica de Tapachula, Carretera a Puerto Madero km 24+300, Tapachula-Puerto Madero (San Benito), 30830 Puerto Madero, Chiapas, M\'exico.}
\date{\today}
\begin{abstract}
  Understanding the properties of Au$_{10}$ clusters entails identifying the lowest energy structure at cold and warm temperatures. While functional materials operate at finite temperatures, energy computations using density functional theory are typically performed at zero temperature, resulting in unexplored properties. Our study undertook an exploration of the potential and free energy surface of the neutral Au$_{10}$ nanocluster at finite temperatures by employing a genetic algorithm combined with density functional theory and nanothermodynamics. We computed the thermal population and infrared Boltzmann spectrum at a finite temperature, aligning the results with validated experimental data. Furthermore, the bonding of Au atoms in lowest energy structure was analyzed employing the quantum theory of atoms in molecules (QTAIM). The Zero-Order Regular Approximation (ZORA) gave consideration to relativistic effects, and dispersion was incorporated using Grimme’s dispersion D3BJ with Becke-Johnson damping. Moreover, nanothermodynamics was utilized to account for temperature contributions. Our findings highlight the preference of small Au$_{10}$ clusters for a planar shape. Despite this, we also consider temperature, relativistic effects, and dispersion from a theoretical point of view. So, the transition from 2D to 3D structures occurs at a larger size. In this work, it is established that the planar lowest energy structure computed with DFT differs from the lowest energy structure obtained with DLPN0-CCSD(T) methodology. The computed thermal population strongly supports the dominance of the 2D elongated hexagon configuration within a temperature range of 50 to 800 K. Importantly, at a temperature of 100 K, the calculated IR Boltzmann spectrum aligns with the experimental IR spectrum. Lastly, the chemical bonding analysis on the lowest energy structure indicates a closed-shell Au-Au interaction with a weak or partially covalent character.
\end{abstract}
\pacs{61.46.-w,65.40.gd,65.,65.80.-g,67.25.bd,71.15.-m,71.15.Mb,74.20.Pq,74.25.Bt,74.25.Gz,74.25.Kc}
\keywords{small gold clusters, Au$_{10}$, density functional theory, temperature, Boltzmann probabilities, Gibbs free energy, entropy, enthalpy, nanothermodynamics, thermochemistry, statistical thermodynamics,  genetic algorithm, DLPNO-CCSD(), the quantum theory of atoms in molecules (QTAIM), DFT, Global minimum}
\maketitle
\section{Introduction}
Transition metal nanoclusters have been extensively researched for their potential applications in catalysis,~\cite{Chaves,https://doi.org/10.1002/anie.200901185,HENRY1998231} photoluminescence,~\cite{doi:10.1021/nl1026716} photonics,~\cite{Barnes2003} magnetism,~\cite{doi:10.1021/nl1026716} chirality,~\cite{doi:10.1021/jp101764q,molecules26185710} and the design of new materials~\cite{doi:10.1063/1.3054185,Majid2020,Buelna-Garcia21}.  The transition metal atoms have incompletely filled {\bf{d}} orbitals that allow them to easily donate and accept electrons from other ions~\cite{DUTTA2023113804,MA2019684}. Gold is the most electronegative metallic atom~\cite{KUHN1998114} and is chemically inert when it is in bulk form~\cite{https://doi.org/10.1002/tcr.10053,SUN2015485,YADAV2020113014,Hammer1995}. However, at nanoscale size, when gold forms small clusters, it is not noble, nonreactive material~\cite{Ghiringhelli_2013}; Thus, the Au clusters find applications in many fields of chemistry, physics, and nanomaterials~\cite{D1CP04440K}. Particularly, gold nanoclusters with diameters below 10~nm~\cite{https://doi.org/10.1002/tcr.10053} are highly catalytically active due to quantum size effects~\cite{SUN2015485,doi:10.1126/science.281.5383.1647}, high oxidation states~\cite{doi:10.1021/ja026998a,doi:10.1021/jacs.5b11768}, and low coordination atoms~\cite{doi:10.1021/ar400068w}, e.g. supported gold nanoclusters catalyze CO oxidation at low temperatures~\cite{YADAV2020113014,Schwank1983} and sub-nanostructured Au species with sizes of 0.9 nm substrates find application on surface-enhanced Raman spectroscopy (SERS)~\cite{LOPEZVAZQUEZ2023116617}. The lowest energy geometries of neutral cluster have been studied many times~\cite{10.1063/1.2179419,D1CP04440K}, both experimental and theoretical point of view, however, is not clear in which cluster size the transition from 2D to 3D take place~\cite{YADAV2020113014,doi:10.1021/jp501701f,D1CP04440K}. 
Bravo-P\'erez  et al.~\cite{Bravo1999} investigated employing the density functional theory (DFT), the neutral Au$_{\mathrm{n}}$ (n=3,7) clusters and found the lowest energy structures are planar. From the experimental point of view, there is evidence that small Au clusters up to n=7 are planar~\cite{10.1063/1.1445121}.

Employing scalar-relativistic pseudopotentials for valence electrons of gold atoms, and under the scheme of DFT, the lowest energy structure of Au$_{\mathrm{n}}$( n=2-10) anionic and neutral clusters were located, they found 3D structures~\cite{PhysRevB.62.R2287}. Furthermore, past research studies investigated the lowest-energy structures of Au$_{\mathrm{n}}$ n=2-20 clusters computed by employing a genetic algorithm coupled to DFT and empirical potential; The authors found that the lowest energy structures are planar and dominated by {\bf{s}} electrons. Moreover, the 2D to 3D transitions take place at Au atomic clusters with seven atoms~\cite{PhysRevB.66.035418}.

For neutral Au clusters, previous high-level \emph{ab initio} calculations show that a structural transition from 2D planar to three-dimensional 3D structures occurs within the size range n=7-10.~\cite{PhysRevLett.89.033401,doi:10.1021/ja040197l,10.1063/1.2150814,10.1063/1.2352755,doi:10.1021/jp505776d} Moreover, the transition could be in 12~\cite{sym14081665} up to 15~\cite{10.1063/1.2434779}.  Assadollahzadeh et al.~\cite{10.1063/1.3204488} reported a systematic search for global lowest energy structures of Au$_{\mathrm{n}}$ n=2-20 clusters employing a genetic algorithm coupled to DFT with a relativistic pseudopotential. They found planar structures up to cluster size Au${10}$. Sarosi et al.\cite{Sarosi2013} reported Au planar putative lowest energy structure employing DFT at B3PW91/def2-TZVPD level of theory. Recently, Pham Vu Nhat et al.~\cite{D1CP04440K} suggest that the 2D to 3D transitions for Au clusters occur at the size Au$_{10}$. However, previous studies on the structure, stability, and bonding in Au$_{10}$ clusters reported the putative global minimum as three dimensional structure at the MP2 level of theory~\cite{DAVID201264}. The 2D to 3D transition in neutral Au clusters is dependent on the level of theory employed~\cite{Sarosi2013} and, with the aim to establish a close-real value of energies, we suggest that they must be computed at a single point CCSD(T) level of theory~\cite{molecules26133953,molecules26185710}. Moreover, the chemistry and physics of gold is strongly dominated by relativistic effects~\cite{https://doi.org/10.1002/hc.10093} e.g. Au-Au bonding distance is shorted due to relativistic effects and binding mode~\cite{doi:10.1021/acs.inorgchem.7b02434}. Relativistic effects have been applied to other systems, Guajardo-Maturana et al. optimized the Ti@Td-C28 structure using scalar relativistic DFT methods with the ADF code~\cite{GUAJARDOMATURANA2024122243}. Previous studies on the structure of small, neutral gold clusters employed the scaled zeroth-order regular approximation to the Dirac equation (ZORA) relativistic correction.~\cite{Ghiringhelli_2013,10.1063/1.467943}, Romaniello et al. ~\cite{10.1063/1.1884985} included relativistic effects for the calculation of linear response properties of metals, particularly the Au atom. Soto et al.~\cite{C3RA46463F} employed ZORA approximation to study the gold cluster cation Au$_7$ they found a putative lowest energy structure 3D. Due to relativistic effects, the preference for the planar shape of neutral gold clusters continues up to 11 atoms~\cite{doi:10.1021/jp505776d,doi:10.1021/cr00085a006,doi:10.1021/jp505776d,10.1063/1.3204488}. Furthermore, Van der Waals interactions (VdW) are critical for predicting the stability of molecules~\cite{LUNAVALENZUELA2021102024,PhysRevMaterials.3.016002,PhysRevLett.91.126402} and can change the energetic ordering of isomers~\cite{ma14010112,molecules26133953,molecules26185710,10.3389/fchem.2022.841964,LUNAVALENZUELA2021102024}. Previous study on small gold clusters in the Au$_{12-14}$ size range found that take into account VdW interactions in the energy calculations, the 2D-3D transition size is with 13 atoms~\cite{Kinaci2016}. Besides relativistic effects and VdW interactions, temperature is a critical factor in all molecular systems, particularly transition-metal nanoclusters that exhibit catalytic activity at low and high temperatures, specifically gold nanoclusters, which have recently received attention because of the discovery of their catalytic activity at low temperature~\cite{Ghiringhelli_2013}. Experiments are performed at finite temperature~\cite{Ghiringhelli_2013,10.1063/1.5132332,10.3389/fchem.2020.00757}, and in realistic conditions materials works at finite temperatures~\cite{10.3389/fchem.2020.00757,10.3389/fchem.2022.841964,molecules26185710,molecules26133953}. However, DFT is developed for zero temperature and ground state computations, e.g.~\cite{ https://doi.org/10.1002/slct.201600525}. In the mid-1960s, Mermin et al. presented a study of thermal properties in inhomogeneous electron gas~\cite{PhysRev.137.A1441}; other previous works included temperature in DFT~\cite{STOITSOV1988121,10.1007/978-3-319-04912-0_2,De_Leon2023-hu}.

Recently, DFT has developed to a finite temperature~\cite{https://doi.org/10.1002/qua.25797,GONIS201886,PhysRevB.82.205120}. Nowdays, temperature effects can be considered by two methods:
\emph{ab initio} molecular dynamics in which temperature is controlled by a thermostat~\cite{C6NR06383G,Martinez-Guajardo2015}. Second, employing quantum statistical mechanics, in which the vibrational
modes or phonon spectrum are needed to compute the molecular partition function that contains
all thermodynamic information~\cite{D1SC00621E,molecules26185710,molecules26133953,https://doi.org/10.1002/chem.201200497,10.1063/5.0061187}. A change in energetic separation among isomer distribution
and dynamic structural rearrangements are the first effects of temperature~\cite{molecules26185710,PhysRevMaterials.3.016002,C1FD00027F,Ghiringhelli_2013}. Particularly, in small gold clusters, the
transitions from 2D to 3D structure depend on temperature. However, few studies have examined
the gold isomers' stability at finite temperature~\cite{C1FD00027F}. Goldsmith et al. studied
neutral small gold clusters at finite temperatures by replica-exchange \emph{ab initio} molecular
dynamics~\cite{PhysRevMaterials.3.016002}. Previous study on neutral Au$_{12}$ predicts dynamical
coexistence of multiple planar and non-planar at room temperature~\cite{PhysRevMaterials.3.016002,PhysRevB.81.174205}. Ghiringhelli et al. studied the structure of small, neutral gold clusters
Au$_3$, Au$_4$, and Au$_7$ and their infrared spectra~\cite {Ghiringhelli_2013}. Previous work
observed the IR spectra in small neutral gold clusters containing up to 8 atoms and compared them
with computed IR spectra~\cite{doi:10.1126/science.1161166}.  In previous research on chemical
bonding in Au clusters, Rodr\'iguez et al.~\cite{RODRIGUEZ2016287} used the QTAIM scheme to
calculate atomic properties and determine the chemical bonding at the Au-gold-Thiol interface.
Recently, Chebotaev et al.~\cite{CHEBOTAEV2023111323} studied the interactions between pterin and
gold clusters using QTAIM theory. Additionally, Zubarev et al.\cite{doi:10.1021/jp808103t} used the
AdNDP method to study chemical bonding in Au$_{20}$.

In this paper, based on quantum statistical mechanics and DFT,
we computed the Gibbs free energy of each Au$_{10}$ isomer and
then calculated the thermal population or so-called Boltzmann population,
a finite temperature. We computed the Boltzmann-IR spectra temperature
dependent based on the thermal population and employed the Boltzmann weights
scheme~\cite{10.3389/fchem.2022.841964,molecules26185710,ma14010112,molecules26133953,doi:10.1098/rsos.230908} To achieve that, we explored the potential
and free energy surfaces of the Au$_{10}$ cluster with a genetic algorithm~\cite{molecules29143374} coupled to DFT employing
ORCA code~\cite{https://doi.org/10.1002/wcms.1606}. The local optimizations
take into account the ZORA relativistic approximation and the Grimme dispersion
D3BJ with Becke-Johnson damping~\cite{https://doi.org/10.1002/jcc.21759} as it
is implemented in Orca code. We study the energetic ordering of low-energy
structures computed at the DFT level against single-point energy calculations
at the DLPNO-CCSD(T) level of theory.
The remainder of the manuscript is
organized as follows. Section 2 provides a brief overview of the theory and
computational details. The results and discussion are presented in Section 3.
We discuss the low-energy structures that take into account ZORA and DBJ3
dispersion, the energetic ordering of isomers at the  DLPNO-CCSD(T) level of
theory, the IR spectra temperature dependent, and the thermal population.
Conclusions are given in Section 4.\par
\section{Theoretical methods and computational details} 
\subsection{Exploration algorithm of the energy landscape of Au$_{10}$ nanocluster}
The search for the lowest energy structure and low-energy structures is a complicated task due to several factors. Firstly, the exploration of the potential energy surface should be systematic and unbiased~\cite{Baletto}. Secondly, the number of possible combinations of atomic arrangements grows exponentially with the number of atoms, leading to a combinatorial explosion problem~\cite{doi:10.1073/pnas.150237097,10.3389/fchem.2022.841964}. Thirdly, the computation of the total energy requires quantum mechanical DLPNO-CCSD(T) level of theory to compute a realistic value of energy. Lastly, sampling a large region of the configuration space is necessary~\cite{Truhlar}.

Several methodos and theoretical studies have developed to search the lowest energy structures~\cite{10.1063/5.0212867}. The design and use of algorithms like simulated annealing,~\cite{kirkpatrick, metropolis, xiang, yang, vlachos, granville} kick method,~\cite{Saunders,Saunders2}
Gradient Embedded Genetic Algorithm (GEGA),~\cite{alexa, alexandrova, alexan} basin hopping,~\cite{harding, wales} among others~\cite{doi:10.1098/rsos.230908,doi:10.1098/rsos.230908,D2CP05188E,doi:10.1021/acs.inorgchem.2c01179,doi:10.1021/jp4099045} helped to explore the potential energy surface. Previous work employed genetic algorithms~\cite{Guo,Dong,Mondal,Ravell,Grande-Aztatzi,RODRIGUEZKESSLER2024122062,doi:10.1142/S2737416524500078} and kick meethodology~\cite{Sudip,Cui,Vargas-Caamal2,Vargas-Caamal,Cui2,Vargas-Caamal2015,Florez,Ravell,https://doi.org/10.1002/chem.201304685,RODRIGUEZKESSLER2024122062} coupled to density functional theory with the aim to explore the potential energy surface.  In this paper, our computational procedure employs a genetic algorithm coupled to Orca code and implemented in \emph{GALGOSON} code~\cite{molecules26133953,ma14010112,Buelna-Garcia21}. The methodology employ a three-step search strategy, the optimization in this first stage was at the PBE0~\cite{Adamo}
and LANL2DZ~\cite{10.1063/1.448975} level of theory. In previous studies, PBEO was employed to compute the density of states that yielded good agreement with the observed spectrum for anion Au$_{19}$.~\cite{doi:10.1021/ja102145g}. The LANL2DZ basis set is employed for transition metals due to its low computational cost~\cite{10.1063/1.448799,molecules26185710} and in second step structures lying in the ten kcal/mol found in the previous step were optimized employing the PBE0 functional and utilizes the basis SARC-ZORA-TZVP~\cite{doi:10.1021/ct800047t} joint with the auxiliary basis SARC/J~\cite{doi:10.1021/ct900090f} and employ the AutoAux generation procedure~\cite{doi:10.1021/acs.jctc.6b01041}. The dispersion is considered by employing the atom-pairwise dispersion correction with the Becke-Johnson damping scheme (D3BJ)~\cite{https://doi.org/10.1002/jcc.21759}. The relativistic effects are considered through SARC-ZORA-TZVP. Additionally, we ensure that each isomer is a true local low-energy structure through the vibrational mode of each isomer, confirming that the lowest vibrational mode of each isomer is positive. Third-step single-point calculations at the DLPNO-CCSD(T) level were performed using the ORCA program suite with TightPNO settings~\cite{doi:10.1021/acs.jpca.9b05734} 
\subsection{Thermochemistry properties}
The partition function, a crucial tool in statistical thermodynamics, is the key to understanding and computing all the thermodynamic properties of an ensemble of atoms at thermodynamic equilibrium~\cite{Dzib,mcquarrie1975statistical,MENDOZAWILSON2020112912,doi:10.1142/S2737416524500078}. Its significance is evident in the Equation~\ref{partition}, where the partition function Q is displayed.
\begin{equation}
\displaystyle
Q(T)=\sum_{i}g_i~e^{{-\Delta{E_i}}/{K_BT}}
\label{partition}
\end{equation}
In Equation.~\ref{partition}, the $g_i$ is the degeneracy, and $k_{\textup{B}}$ is Boltzmann constant, $T$ is the temperature absolute, and ${-\Delta{E_i}}$ is the total energy of a molecule~\cite{Dzib,mcquarrie1975statistical}. To calculate Q precisely, we need to account for the coupling of internal modes. One way to simplify this is by using the Born-Oppenheimer approximation, which assumes that the electron movement is much faster than that of the nuclei. This approximation treats the molecular wave function as a product of the electronic and nuclear wavefunctions: $\psi=\psi_e\psi_n$. The vibrational and rotational degrees of freedom can be closely linked, and their separation is referred to as the Harmonic Oscillator Rigid Rotor approximation. In this approximation, vibration is characterized using a harmonic oscillator model and rotations are described using the rigid rotor model. When the Born-Oppenheimer and Harmonic Oscillator Rigid Rotor approximations are applied, the molecule's total energy is the sum of electronic, translational, vibrational, and rotational energies. As a result, the partition function can be expressed as a product of the associated contributions~\cite{hill1986introduction,Buelna-Garcia21} The starting  point to compute the thermochemical data is the partition function given in Equation~\ref{partition} and  it can be decomposed as contributions from rotational translation, vibration and electronic. Therefore, the total partition function of a molecule in its ground state is computed as
the product of all partial partition functions, then  we can re-write Equation~\ref{partition} as a product of partial partition functions~\cite{hill1986introduction,Dzib, ma14010112} as given in Equation~\ref{p2}.
\begin{equation}
\displaystyle
\mathrm{Q}=\mathrm{q_{trans}~q_{rot}~q_{vib}~q_{elec}}
\label{p2}
\end{equation}
The contributions of electronic, translational, vibrational, and rotational partition functions to the canonical partition function $Q$, given in Equation~\ref{p2} are given in Ref.~\cite{ma14010112}
All those Equations are calculated at a standard pressure of 1 atm and at a specific temperature using ideal gas statistical mechanics. These equations are equivalent to those found in the reference (Ref.~\cite{Dzib}) and in standard thermodynamics texts~\cite{mcquarrie1975statistical,hill1986introduction}.

In this study, q$_{\mathrm{trans}}$ is calculated at a finite temperature and is used to determine the translational entropy. The rotational contribution depends on the moments of inertia and the rotational symmetry number~\cite{molecules26133953}. The Rigid Rotor Harmonic Oscillator approximation is used to separate rotational and vibrational modes~\cite{ma14010112}. Previous works have computed the difference in rotational entropy with and without considering symmetries~\cite{molecules26133953}.

To account for the breakdown of the harmonic oscillator approximation at low frequencies, entropic contributions to the free energies are computed using the Quasi-RRHO approach of Grimme~\cite{https://doi.org/10.1002/chem.201200497}. The electronic partition function, q$_{\mathrm{elec}}$, is represented by q$_{\mathrm{elec}}=\omega_0$.

The total energy contribution of a molecule at a finite temperature is the sum of electronic, translational, vibrational, and rotational energies. Equations~\ref{e1} to~\ref{ag22} are applied to calculate the internal energy (U), enthalpy (H), and Gibbs free energy (G) of the Au$_{10}$ cluster at a finite temperature. The rotational contribution to the entropy is calculated using the expressions given by Herzberg~\cite{herzberg1945infrared}.
\begin{equation}
\displaystyle
\mathcal{U}_0=\mathcal{E}_0+ZPE\\
\label{e1}
\end{equation}
\begin{equation}
\displaystyle
U_T=\mathcal{U}_0+\mathrm{(E_{Rot}+E_{Trans}+E_{vib}+E_{elect})}\\
\label{e2}
\end{equation}
\begin{equation}
  \centering
  \displaystyle
   H=U+nRT\label{ag11}
\end{equation}
\begin{equation}
  \centering 
  \displaystyle
  \Delta G=\Delta H-\Delta ST\label{ag22}
\end{equation}
In Equations above, ZPE is the zero-point energy correction  $\mathcal{E}_0$ is the electronic energy,
and $\mathrm{E_{Rot}+E_{Trans}+E_{Vib}+E_{elect}}$ are the contributions to energy due to
translation, rotation, and vibration as function of temperature, respectivly.
In Equations~\ref{ag11}~and~\ref{ag22}, R is the ideal gas constant and $n$ is the amount of substance, and T is the absolute temperature.
\subsection{Thermal population}
The mesured properties in a molecule are statistical averages over the ensemble of geometrical conformations or isomers accessible to the cluster~\cite{Teague}. In an Boltzmann ensemble at thermal equilibrium  of Au$_{10}$ clusters we computed the thermal population ~\cite{D1SC00621E,molecules26185710,Truhlar,MENDOZAWILSON2020112912,Dzib,Schebarchov,Grigoryan}
employing Equation~\ref{boltzman1} 
\begin{equation}
\centering 
\displaystyle
P_i(T)=\frac{e^{-\beta \Delta G^{k}}}{\sum e^{-\beta \Delta G^{k}}}\label{boltzman1},
\end{equation}
where $\beta=1/k_{\textup{B}}T$, and $k_{\textup{B}}$ is the Boltzmann constant, $T$ is the temperature, and $\Delta G^{k}$ is the Gibbs free energy of the $k^{th}$ isomer.
Equation~\ref{boltzman1} is restricted to  the sum of all
thermal populations at finite temperature, $P_i(T)$ is equal to 1, according
to Equation~\ref{bol2}
\begin{equation}
\centering 
\displaystyle
\sum_i P_i(T)=1\label{bol2},
\end{equation}
In this paper, the Boltzmann  weighted IR spectrum (IR$_{Bolt}$) at at finite temperature  is given by Equation~\ref{vcd}
\begin{equation}
\displaystyle
IR_{Bolt}=\sum_i^{n}IR_{i}\times P_i(T)
\label{vcd}
\end{equation}
Where $n$ is the total number of cluster in the ensemble, IR$_{i}$ is the IR of the $i^{th}$  isomer at T=0, and  P$_i$(T) is the thermal population of the $i$ isomer given by Equation~\ref{boltzman1}.
\subsection{The chemical bonding analysis}
\begin{figure}[htbp!] 
  \centering
    \includegraphics[scale=0.40]{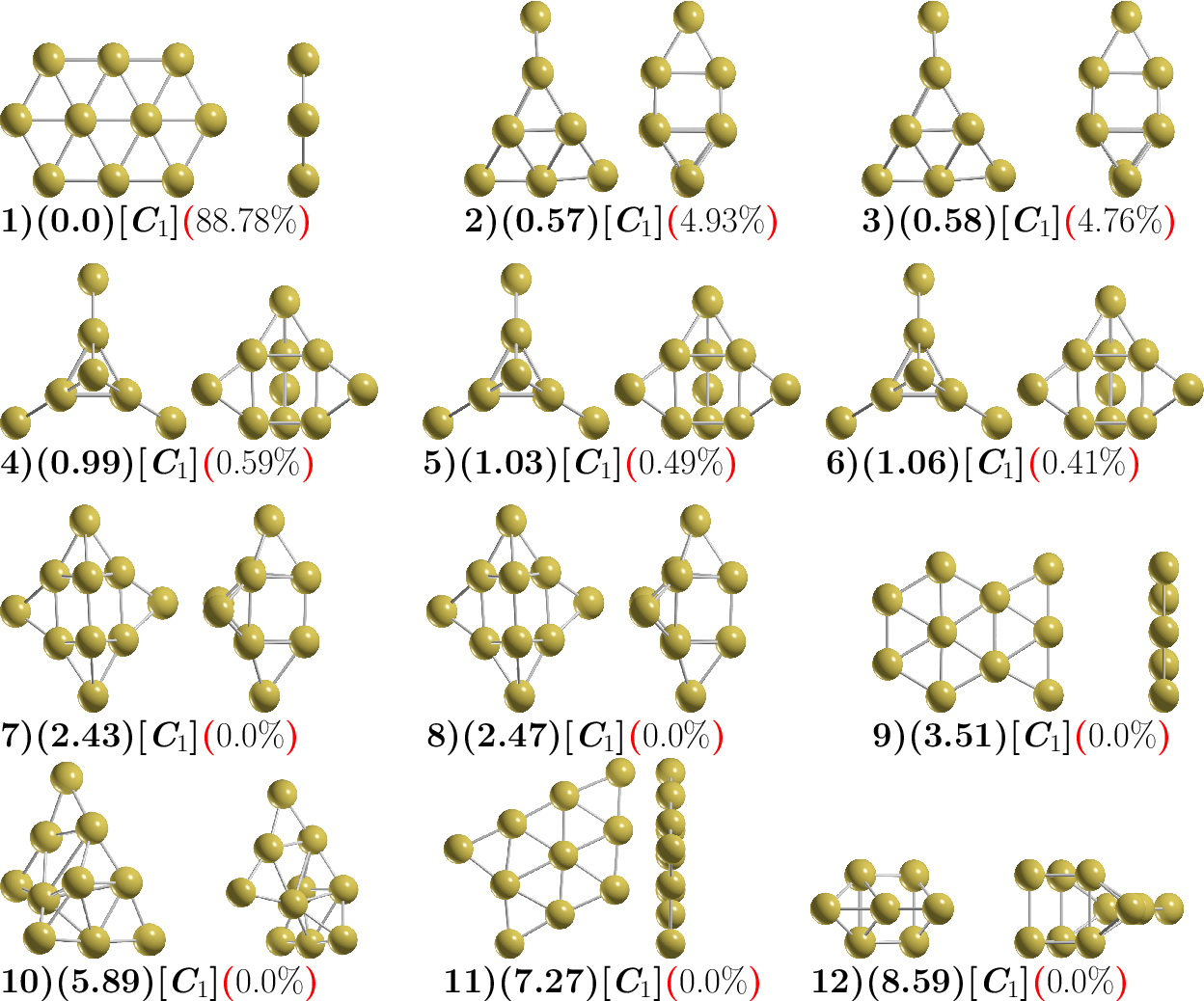}
    \caption{(Color online) The optimized geometries of neutral Au$_{10}$ cluster at PBE0 ZORA SARC-ZORA-TZVP  SARC/J D3BJ level of theory. The most important energy isomers are show in two orientations. In round parenthesis, relative Gibbs free energies in kcal/mol. The point group symmetry is displayed in square brackets, while in red round parenthesis the thermal population at 100 K. The yellow-colored spheres represent the Au atoms.}
  \label{georo}    
\end{figure}
We carry out an analysis of the electron density, based on the quantum theory of atoms in molecules (QTAIM)~\cite{doi:10.1021/cr00005a013,KUMAR2016,Guevara-Vela2024} for determining the chemical bonding at the lowest energy structure of Au$_{10}$ cluster employing  the MultiWFN program~\cite{https://doi.org/10.1002/jcc.22885}. The topology of the electron density is based on the critical point (CP) concept. The CP is deﬁned as the spatial point where the ﬁrst derivative of the electron density vanishes. The bond critical point (BCP) is found between two nuclear critical points (NCP) and indicates the presence of a chemical bond.
\section{Results and Discussion}
\subsection{The lowest energy structures and energetics}
The final optimized geometries depicted in Figure ~\ref{georo} are reported at PBE0~\cite{Adamo}-D3BJ~\cite{https://doi.org/10.1002/jcc.21759} and ZORA~\cite{Ghiringhelli_2013,10.1063/1.467943} SARC-ZORA-TZVP~\cite{doi:10.1021/ct800047t}  SARC/J~\cite{doi:10.1021/ct900090f} and employing AutoAux generation procedure~\cite{doi:10.1021/acs.jctc.6b01041,doi:10.1021/acs.jctc.3c00670}, level of theory. All calculations were performed using  the version 5.0 of the ORCA quantum chemistry program suite~\cite{doi:10.1021/acs.jpca.9b05734}.
Figure~\ref{georo} shows the 12 most stable structures obtained for the Au$_{10}$ cluster with relative Gibbs free energies computed at 100 K, 1 atm, and lying in the 0-8.6 kcal/mol energy range. The putative global minimum is taken as a reference structure for comparison.
The planar elongated hexagon depicted in Figure~\ref{georo}(1) is the lowest energy structure, and it is a planar structure composed of triangular units with symmetry \emph{C$_1$}. The computed average optimized Au-Au bond length is 2.56~\AA~that is slightly higher than the experimental bond distance of 2.4715~\AA~obtained for the Au dimer from the high-resolution rotational spectroscopy~\cite{SIMARD1990310,10.1063/1.2179419}. All the structures displayed in Figure~\ref{georo} are optimized low-energy structures without any imaginary vibrational frequencies, so there are no transition states. The planar elongated hexagon depicted in Figure~\ref{georo}(1) has also been reported earlier as a putative global minimum for different authors~\cite {Sarosi2013} employing DFT but without considering temperature. In agreement with previous works, the elongated hexagon is the lowest energy structure at 100 K~\cite{PhysRevMaterials.3.016002}. The second structure with higher Gibbs free energy lies at 0.57 kcal/mol above the putative global minimum and shown in Figure~\ref{georo}(2), its thermal population is 4.93\%. It consists of a trigonal prism as the main central structure, which is edge-capped by three atoms and one atom-capped on one side, composed of four atoms of the central structure. The third structure's higher Gibbs free energy lies at 0.58 kcal/mol, its thermal population is 4.76\%, and it is slightly distorted compared with the previous structure. The following structure is located 0.99 kcal/mol above the putative global minimum and is shown in Figure~\ref{georo}(4); its thermal population is 0.56\%. It is composed of a trigonal prism as the main central structure, which is edge-capped by three atoms with one atom capped on the top. It is composed of three atoms of the main central structure. This structure is found in Sn-type clusters as low-energy structures~\cite {doi:10.1021/jp8030754}. Our calculated structure, Figure~\ref{georo}(4), is in agreement with reported previously as the low-lying energy structures~\cite{D1CP04440K}. The following two isomers located 1.03 and 1.06 kcal/mol energy above the putative global minimum and shown in Figure~\ref{georo}(5), and Figure~\ref{georo}(6) are similar to the structure depicted in Figure~\ref{georo}(4). Those isomers are slight distortions, with a thermal population of 0.59, 0.48, and 0.41{\%}, respectively.
The following two isomers shown in Figure~\ref{georo}(7,8) are also distorted capped trigonal prism with 2.43 and 2.47 kcal/mol, respectively above the putative global minimum, and with a thermal population of zero.  Nhat et al.~\cite{D1CP04440K} used  PNO-LCCSD(T)-F12/aug-cc-pVTZ-PP level of theory, plus ZPE energy correction based on PBE/cc-pVTZ-PP optimized geometries to explore the potential energy surface of the Au$_{10}$ cluster and found the capped trigonal prism laying to 1.6 kcal/mol above the trigonal prism edge-capped. Now we examine the structures located at 3.51 and 7.27 kcal/mol lying in energy above the lowest energy planar elongated hexagon and displayed in Figures~\ref{georo}(9,11). In previous studies, those isomers were found to be competing planar structures at 300 K~\cite{PhysRevMaterials.3.016002}. The structure displayed in Figure~\ref{georo}(9) is reported as a local minimum employing PNO-LCCSD(T)-F12/aug-cc-pVTZ-PP level of theory\cite{D1CP04440K} and located at 3.3 kcal/mol above the putative global minimum. The distorted structure displayed in Figure~\ref{georo}(10) and located at 5.89 kcal$\cdot$mol$^{-1}$ is a folded flake structure not previously reported.
 To explore spin effects in determining the lowest energy structure of the Au$_{10}$ cluster, we investigate multiplicities up to the triplet state. All neutral Au$_{10}$ clusters optimized with spin multiplicities of 1 were taken as initial structures for optimization to higher spin states up to triplet; for all these geometries, optimization without any symmetry restriction was performed considering spin multiplicities 3. Our computed structures indicate that the low-lying energy structure in the triplet state is located at 30 kcal/mol above the putative global minimum in the singlet state. The second low-energy structure in the triplet state is located at 33 kcal/mol above the lowest-energy structure in the singlet state. Thus, higher spin states (up to triplet) are not energetically favored.
\subsection{The chemical bonding analysis}
To shed light in the bonding of the Au$_{10}$ cluster we carry out the chemical bonding analysis on the 2D enloganted hexagon structure depicted in Figure~\ref{georo}(1).
\subsubsection{Quantum theory of atoms in molecules (QTAIM) analysis}
\begin{figure}[htbp!]
  \centering
    \includegraphics[scale=0.5]{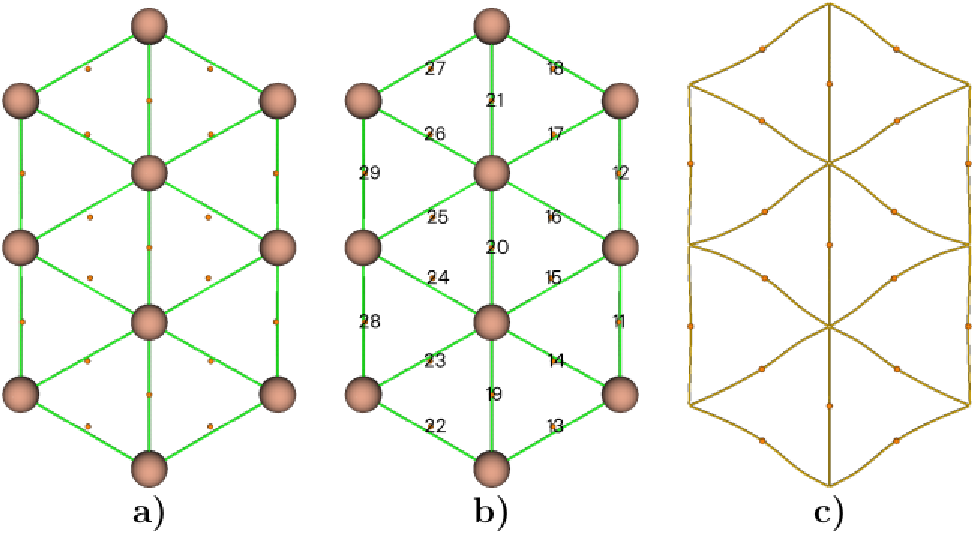}
    \caption{(Color online) Molecular graphs for the lowest energy structure  Au$_{10}$ cluster. a) BCP are in red-yellow color small spheres. b) Index of the BCP, c) The bond paths (BP), they are lines of maximum density that connects a pair of nuclear critical points. The BP's are used as an indicator of chemical bonding~\cite{https://doi.org/10.1002/anie.200805751} between a pair of Au atoms.  The Au atoms are depicted as red-brown spheres.}
     \label{qtaim} 
\end{figure}
The electron density values computed at the BCP provide information about the nature of the interaction between the two Au atoms~\cite{CHEBOTAEV2023111323}. The total number of critical points of each type we found for the lowest energy structure of the Au$_{10}$ cluster are 10 NCP, 19 BCP, and 10 Ring Critical Points.
Figure~\ref{qtaim} displays a) the BCP points, b) the index of BCP, and c) the BP paths. We found 39 CP points and all of them fulﬁll Poincare-Hopf relationship. According to our computed CP displayed in Figure~\ref{qtaim} a), there are 19 BCP that indicate there are chemical bonds among Au-Au atoms. Figure~\ref{qtaim} c) shows the BP indicative of chemical bonding. To shed light on the chemical bonding in each BCP, we evaluated those QTAIM parameters: the electronic density ($\rho$), the Laplacian of electron density ($\nabla^2 \rho $), and the energy density H(r), Displayed in Table 1. Analyzing the data of Table 1, the Au-Au average bond distance between Au-Au atoms is 2.7034 $\textup{\r{A}}$, which is in agreement with the bond distance reported for gold nanoclusters, which are in the range 2.6-2.84 $\textup{\r{A}}$ ~\cite{RODRIGUEZ2016287}. From Table 1, the Laplacian of electron density is positive for all BCPs; the positive value of the Laplacian at the BCP points indicates a depletion of the density~\cite{RODRIGUEZ2016287}. Thus, according to the values displayed in Table 1, the Au-Au bond is non-covalent, in agreement with previous work~~\cite{RODRIGUEZ2016287}. Similar weak non-covalent Pd-Pd interactions have been reported in heterobimetallic (Pd-Pd) or heterobimetallic (Pd-Ir) clusters~\cite{LEPETIT2017150}. Positive Laplacian of electron density and negative energy density values indicate a closed-shell Au-Au interaction with a weak or partially covalent character. 
\begin{table}[htbp!]
\centering
\begin{tabular}{llllc}
  \hline
  \multicolumn{5}{c}{BCP data employing QTAIM theory} \\
  \hline
   Indeces CP & $\rho$ & $\nabla^2 \rho $ &  H(r) & Bond distance \AA  \\
   \hline
   14   &  0.5748&      0.1145 &       -0.1810 &    2.6892\\
19    & 0.7227    &  0.1291    &    -0.2436   &  2.6770 \\
23    & 0.5748    &  0.1145    &    -0.1810   &  2.6892\\
24    & 0.5358    &  0.1100    &    -0.1478   &  2.7173\\
20    & 0.7130    &  0.8501    &    -0.2287   &  2.7304\\
15    & 0.5358    &  0.1100    &    -0.1478   &  2.7173\\
\hline
\end{tabular}
\caption{1) Index of CP as they are displayed in Figure~\ref{qtaim}a. 2) The electronic density ($\rho$). 3) The Laplacian of electron density ($\nabla^2 \rho $). 4) The Energy density H(r). 5) Bond distances.}
\label{qtaim2}
\end{table} 
\subsection{Energetics}
To shed light on the true energy hierarchy of the Au$_{10}$ isomers and compare with DFT energies. Single point energies of each isomer were computed employing the method of DLPNO-CCSD(T) with TightPNO setting corrected with zero-point energy ZPE based on  PBE0 ZORA SARC-ZORA-TZVP  SARC/J D3BJ level of theory of optimized geometries,  and compared with those obtained at the DFT level of theory. Previous studies pointed out that employing different methods to compute energies yields different results~\cite{molecules26185710}, particularly in small gold clusters~\cite{sym14081665}. 
\begin{table}[htbp!]
\centering
\begin{tabular}{ p{2cm} | p{2cm} | p{3cm}}
\hline
\multicolumn{3}{c}{Energy DFT versus DLPNO-CCSD(T) SP energy } \\
\hline
No. of Isomer (Fig. 1) & DFT energy (kcal/mol)  & DLPNO-CCSD(T) SP energy (kcal/mol) \\
\hline
1 & {\framebox{0.0}} & 4.26 \\
2 &  0.57 & 8.44 \\
3 &  0.58 & 9.34 \\
4 &  0.99 & 0.03 \\
5 &  1.03 & 0.53 \\
6 &  1.06 & {\framebox{0.0}}\\
7 &  2.43 & 3.18 \\
8 &  2.47 & 3.73 \\
9 &  3.51 & 3.92 \\
10 & 5.89 & 4.56 \\
11 & 7.27 & 7.11 \\
12 & 8.59 & 15.08 \\
\hline
\end{tabular}
\caption{The first row indicates the number of isomers as per Figure 1. In the second row, we calculated the relative DFT Gibbs free energy using the PBE0-D3BJ and ZORA SARC-ZORA-TZVP SARC/J levels of theory at 100 K. In the third row, we computed the relative single-point energy using domain-based local pair natural orbital coupled-cluster theory (DLPNO-CCSD(T)) with the TightPNO setting.}
\label{ene}
\end{table}
Table~\ref{ene}, in the first row, displays the number of isomers following the label-cluster given in Figure~\ref{georo}. The second row of Table~\ref{ene} displays the relative DFT Gibss free energy computed at PBE0-D3BJ and ZORA SARC-ZORA-TZVP SARC/J level of theory and at 298.15 K. The Third row of Table~\ref{ene} displays single point energy computed at DLPNO-CCSD(T)  level of theory plus ZPE energy and follows the energetic ordering of the second row of  Table~\ref{ene}. At the DFT level of theory, the second row of Table~\ref{ene}) indicate that the lowest energy is the planar elongated hexagon structure depicted in Figure~\ref{georo}a. According to single point energy at the DLPNO-CCSD(T) level, displayed in row third of Table~\ref{ene}, this isomer lies at 4.26 kcal/mol above the putative lowest energy structure. This is composed of a trigonal prism as the main central structure that is edge-capped by four atoms, and it is depicted in Figure~\ref{georo}b and listed isomer label number five in table~\ref{ene}. In summary, Table~\ref{ene} indicates that the lowest energy structure is the 2D elongated hexagon structure computed at the DFT level of theory, while at the  DLPNO-CCSD(T)+ZPE level of theory is the 3D trigonal prism. According to  DLPNO-CCSD(T)  energies, the second isomer, labeled number 3 Table~\ref{ene}, is at 0.03 kcal/mol above the putative global minimum. the next isomer, labeled number 6 in Table~\ref{ene}, is located at 0.53 kcal/mol energy above. The computed thermal population for those three isomers is 42, 40, and 17{\%}, respectively. The thermal population displayed in Figure~\ref{popuG} was calculated using the Gaussian code without ZORA approximation (Appendix A). Au$_{10}$ atomic coordinates in XYZ format can be found in Appendix B.
\section{Thermal Population}
\begin{figure}[htbp!]
 \centering  
 \includegraphics[scale=0.80]{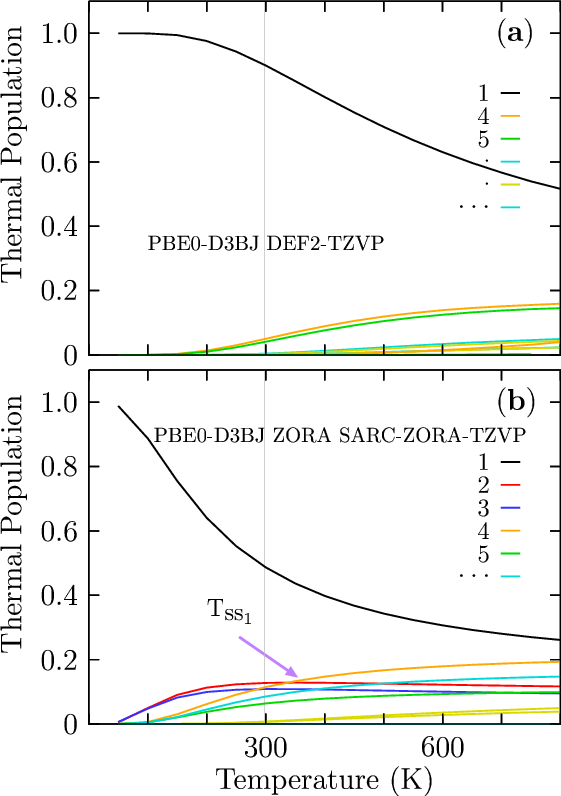}
 \caption{(Color online) Thermal population for all Au$_{10}$ isomers at temperatures ranging from 50 to 800  K. For easy comparison,  in panel (a), we show the thermal population without relativist effects. In panel (b), we show the thermal population where relativist effects are taken through Zero-Order Regular Approximation (ZORA). The melting temperature of gold is 1064 K~\cite{JelenKraje}; thus, our results below this temperature are valid. The thermal population was calculated using the Gaussian code without ZORA (Appendix A). Au$_{10}$ atomic coordinates in XYZ format can be found in Appendix B.}
 \label{lboro}
 \end{figure}
In this section, based on the computed free-energy differences among Au$_{10}$ isomers,
 we calculated the thermal population~\cite{D1SC00621E,molecules26185710,Truhlar,MENDOZAWILSON2020112912,Dzib,Schebarchov,Grigoryan} or Boltzmann distribution at finite temperature and standard pressure, as given in Equation~\ref{boltzman1}. The observed molecular properties are statistical averages over of a collection of molecules;  As consequence, the statistical averages  are crucial for the experiment measurement of thermodynamic quantities~\cite{Palma2015} 
 The separation among isomers by energy differences is a critical energy that determines the transition solid-solid point on the thermal population. Suppose the free energy difference between the putative lowest energy structure and the first low-energy isomer is small ($<$ 1 kcal/mol). The molecular system is related to several solid-solid transition points in that case.
 Conversely, if those free energy differences are more significant ($>$ 1 kcal/mol), the thermal population has no transition solid-solid points. Additionally, free-energy differences are related to fundamental chemical quantities such as binding constants, solubilities, partition coefficients, and adsorption coefficients~\cite{Gunsteren,doi:10.1021/ct500161f,10.1063/1.2730508}. Figure~\ref{lboro}, panel (a), displays the thermal population for each particular  Au$_{10}$ isomer for temperatures ranging from 50 to 800 K. We computed that thermal population at PBE0 Def2-TZVP  def2/J RIJCOSX  level of theory taking into account the D3BJ dispersion correction method. The thermal population of the lowest energy planar elongated hexagon configuration is depicted by a black solid line in Figure~\ref{lboro}, panel (a);  in temperatures ranging from 50 to 150 K, the thermal population is constant. At 150 K, it started to decay exponentially up to room temperature, where it achieved 90\%  probability. Above room temperature, the thermal population decays almost linearly, achieving 50\% at 800 K, as it is displayed in the upper panel of Figure~\ref{lboro}  panel (a). The analysis of these results led to three interesting conclusions in temperatures ranging from 50 to 800 K. 1)  there are no solid-solid transformation points, 2)  the planar elongated hexagon configuration is strongly dominant, and 3) Almost all molecular properties of the Au$_{10}$  cluster are attributed to planar elongated hexagon configuration (89\%). The thermal population of the isomer, edge-capped trigonal prism displayed in Figure~\ref{georo}(4), is depicted by a yellow solid line in Figure~\ref{lboro}, panel (b). At a temperature of 100 K, it starts to increase, achieving 15\% probability at room temperature, and at a temperature of 800 K, it achieves almost 15\% of probability. This low-energy structure  contribute to the molecular properties only 0.59\% at 100 K.  
\section{Infrared spectrum (IR) of {{{Au}$_{10}$}} cluster} 
\begin{figure}[ht!]
 \centering  
 \includegraphics[scale=0.50]{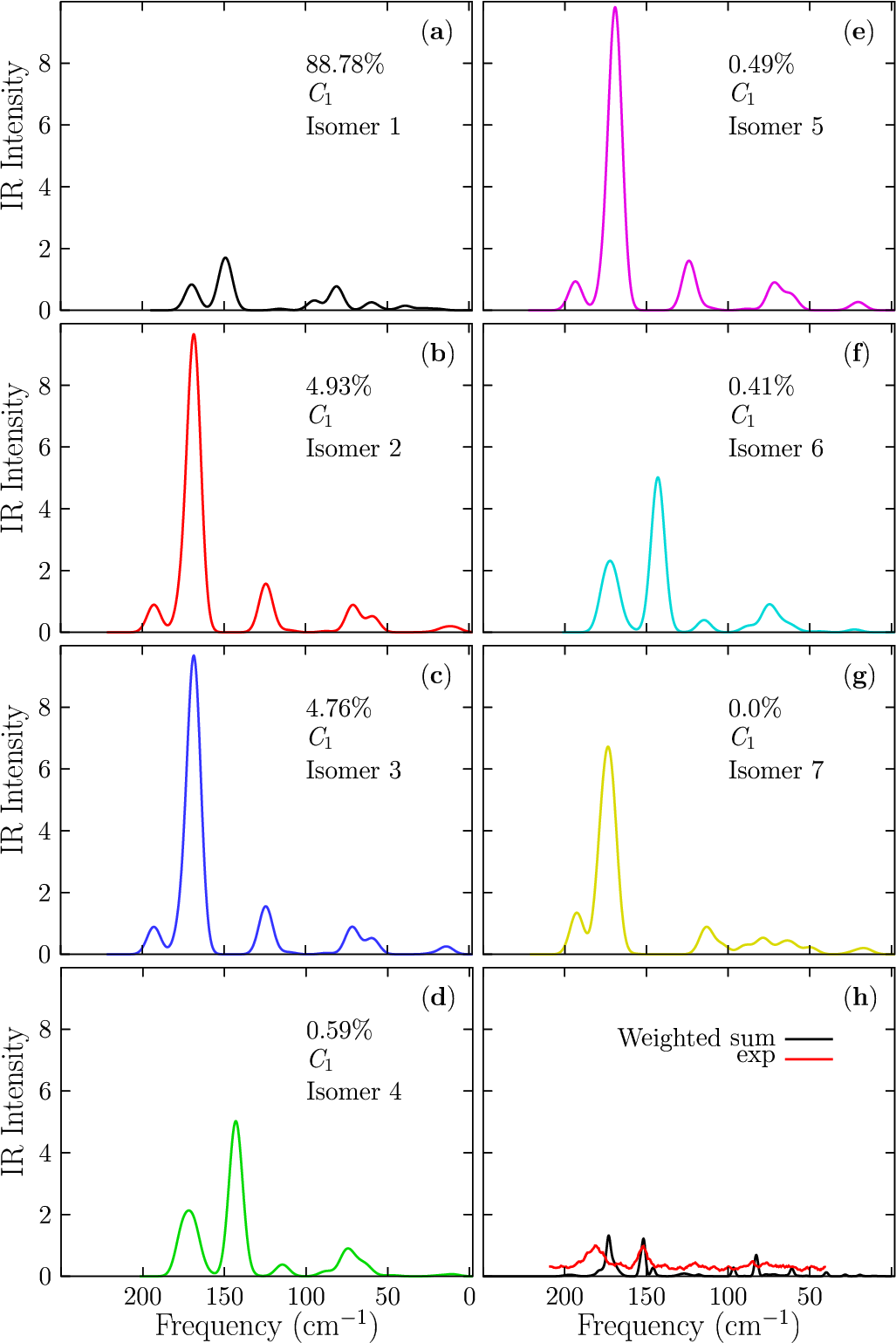}
 \caption{(Color online) Computed and experimental  IR spectra of Au$_{10}$ isomers. (a-g) Computed infrared spectra of Au${10}$ cluster at 0 k and in the range of 50 to 250 cm$^{-1}$  (h) A comparison of the IR experimental and the Boltzmann spectra computed at room temperature and depicted as black solid-line. The IR Boltzmann spectrum is a weighted sum of each isomer’s IR spectra. The red solid line is the experimental infrared spectrum taken from references~\cite{D1CP04440K,PhysRevMaterials.3.016002}}
 \label{ir}
\end{figure}

IR spectra are usually used to identify functional groups and bond chemical information; however, from the experimental point, the IR bands' assignment to vibrational molecular modes can be somewhat difficult and requires \emph{ab initio} calculations. In these computations, the temperature is generally not considered, and discrepancies between experimental IR spectra and calculated IR spectra can result from finite temperature and anharmonic effects. In this study, each spectrum of each isomer was computed using DFT, as implemented in the Gaussian 09 code. Figure~\ref{ir}, panels (a) to (g) display the individual IR spectra that belong to the putative global minima and the six lowest energy structures located in the relative energy range up 0 to 2.43 kcal/mol. In panel (h), we show a comparison of the experimental IR spectrum~\cite{D1CP04440K,PhysRevMaterials.3.016002} and the computed Boltzmann spectrum of Au$_{10}$ cluster. Depicted in black solid-line, we show the Boltzmann weighted spectrum at 298.15 K. The red solid-line is the experimental infrared spectrum of Au$_{10}$ taken from references~\cite{D1CP04440K,PhysRevMaterials.3.016002}
The experimental spectrum displayed in Figure~\ref{ir}, panel (h), shows two peaks located at 150 cm$^{-1}$ and  180 cm$^{-1}$, respectively. The spectrum is almost flat In the range 50 to 150 cm$^{-1}$. Interestingly, the computed IR Boltzmann spectrum is in good agreement with the experimental studies of Goldsmith et al.~\cite{D1CP04440K,PhysRevMaterials.3.016002}. The peak of the IR Boltzmann spectrum, depicted with a black solid line that is located at 150 cm$^{-1}$, shows good agreement with the experimental IR spectrum, as we can see in Figure~\ref{ir} panel (h). Moreover, the intensities of those peaks are similar. The vibrational mode of the peak is located at 150 cm$^{-1}$; it consists of the stretching and oscillation of the two central atoms along the larger axis of the cluster and within the plane. The peak located at 180 cm$^{-1}$ consists of stretching and oscillating the two central atoms along the shorter axis of the cluster and within the plane. Interestingly, the 88.7{\%} of the Boltzmann IR spectrum, depicted in a black solid line and displayed in Figure~\ref{ir} panel (h), comes from the IR spectrum of the putative global minimum. The other 11.3{\%} of the Boltzmann IR spectrum are contributions from the three low-energy structures. 
\section{Conclusions}
In summary, we explored the potential and free energy surface of the Au$_{10}$ cluster using an efficient genetic algorithm written in \emph{Python} and coupled to density functional theory
 through version 5.0 of the ORCA quantum chemistry program suite. Our computations show that the putative lowest energy structure, at 0 K, is the  2D elongated hexagon configuration. If the system's temperature increases, entropic effects start to play, and Gibbs's free energy determines the lowest energy structure. Therefore, at warm temperatures, our computed thermal population indicates that the planar elongated hexagon configuration strongly dominates in the temperature range of 0 to 800 K. We take into account relativistic effects through the ZORA approximation, dispersion through D3BJ Grimme methodology and temperature through nanothermodynamics. Considering dispersion and ZORA approximation, our computed thermal population indicates that planar elongated hexagon configuration strongly dominates in all temperature scales. However, the thermal population without relativistic effects indicates that the planar elongated hexagon configuration strongly dominates all temperatures. As mentioned, the relativistic and temperature effects do not interchange the 2D elongated hexagon configuration with a 3D structure (prism as the main central structure that is edge-capped by four atoms). Additionally, a comparison of the energetic ordering of lowest structures computed at DFT energies and the single point energies employing the domain-based local pair natural orbital coupled-cluster with single, double, and perturbative triples excitation (DLPNO-CCSD(T)) using the ORCA program.
 We found that the putative global minimum energetic ordering computed at the DFT level of theory is not the lowest energy structure at DLPNO-CCSD(T). There is an interchage in the energetic ordering of the isomers computed at DLPNO-CCSD(T) level of theory.
 We computed the IR spectrum dependent on temperature at 100 K, employing the Boltzmann methodology. Our computed Boltzmann IR spectrum is in good agreement with experimental studies realized at 100 K of B. R. Goldsmith et al.when we take into account the energetic ordering of isomers computed at the  DFT PBE0-D3BJ  and ZORA  SARC-ZORA-TZVP  SARC/J. We emphasize that the IR spectrum calculated using the DLPNO-CCSD(T) SP energetic ordering of isomers does not correspond to the experimental IR spectrum. It appears that the 3D energetic global minimum at DLPNO-CCSD(T) doesn't match the experimental one.
Based on the excellent agreement on the computed  IR spectrum versus IR experimental, both at a temperature of 100 K, We can conclude that the 2D elongated hexagon configuration is the absolute global minimum in the 0 to 800 K temperature range.
The analysis of chemical bonding was conducted within the framework of the quantum theory of atoms in molecules. The positive Laplacian of electron density and negative energy density values indicate a closed-shell Au-Au interaction with a weak or partially covalent character.
Shortly, we will focus on calculating the optical spectra in Au$_{10}$ clusters using the Boltzmann scheme.
\section{Acknowledgments}
 F.E.R.-G. thanks Conahcyt-M\'exico for the Ph.D. scholarship 492050.  Computational resources partial for this work were provided by \emph{ACARUS} through the High-Performance Computing Area of the University of Sonora, Sonora, M\'exico. We are also grateful to the computational chemistry laboratory for providing computational resources, \emph{ELBAKYAN}, and \emph{PAKAL} supercomputers of  Polytechnic University of Tapachula. 
\section{Conflicts of Interest} The authors declare no conflict of interest.
\section{Funding} This research received no external funding.
\section{Abbreviations}
The following abbreviations are used in this manuscript:
Density Functional Theory (DFT), Domain-based Local Pair Natural Orbital Coupled-Cluster Theory (DLPNO-CCSD(T)), Zero-Point Energy (ZPE), Global Genetic Algorithm (GALGOSON) quantum theory of atoms in molecules (QTAIM).  
\bibliographystyle{unsrt}
\bibliography{manuscript}
\newpage

\appendix
\onecolumngrid
\section{Appendix A. Boltzmann Probabilities computed at TPSS-GD3/Def2TZVP  and B3PW91-GD3/Def2TZVP with Gaussian code.}
\label{appendix:a}
\begin{figure}[ht!]
\centering  
\includegraphics[scale=0.65]{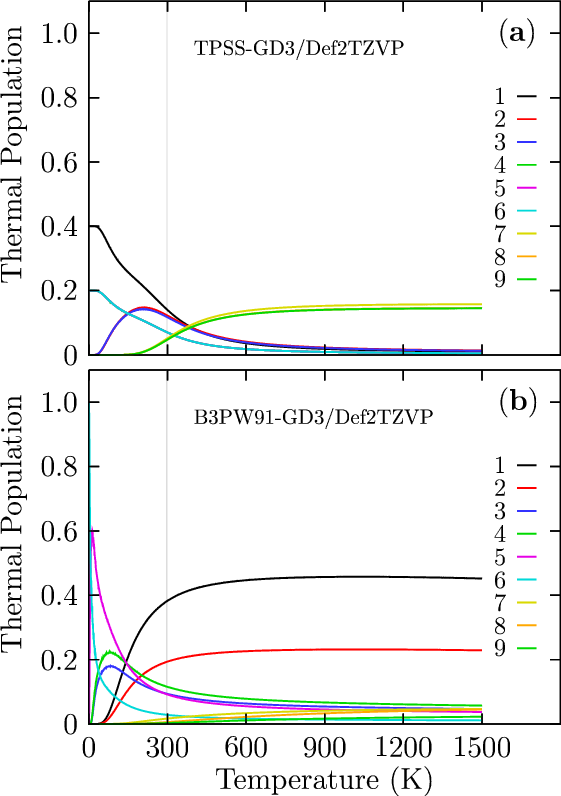}
\caption{(Color online) Thermal population (probability of occurrence) for all isomers Au$_{10}$ at temperatures ranging from 20 to 1500 K computed with  DFT  functionals and basis set: (a) TPSS-GD3/Def2TZVP  and (b) B3PW91-GD3/Def2TZVP.  The melting temperature of gold is 1064 K~\cite{JelenKraje}; thus, our results below this temperature are valid. The thermal population was calculated using the Gaussian code without ZORA approximation (Appendix A). Au$_{10}$ atomic coordinates in XYZ format can be found in Appendix B.} 
\label{popuG}
\end{figure}
\newpage 
\section{Appendix B. XYZ atomic coordiantes}
\label{appendix:b}
\begin{verbatim}
----------------Orca code----------------------------
10  
0.000000000         0001.orco
Au     -2.682054000      2.344828000     -0.000000000
Au      4.042240000      0.000006000     -0.000000000
Au     -2.682080000     -2.344830000     -0.000000000
Au      2.682077000     -2.344826000      0.000000000
Au      0.000001000      2.349466000      0.000000000
Au     -4.042241000      0.000005000     -0.000000000
Au     -1.365241000     -0.000002000      0.000000000
Au      0.000000000     -2.349474000      0.000000000
Au      2.682057000      2.344828000      0.000000000
Au      1.365240000     -0.000001000      0.000000000
10  
0.799240662         0002.orco
Au      0.854742000     -1.608527000     -0.807029000
Au     -0.657247000      1.936475000     -1.291186000
Au     -1.053479000      1.153378000      1.235045000
Au     -1.761483000     -0.451313000     -0.813926000
Au     -1.394093000     -3.145368000     -0.537260000
Au      1.295575000      2.939223000     -2.776782000
Au      1.875284000      0.815873000     -1.291881000
Au     -0.734166000     -1.654581000      1.560167000
Au      1.566606000     -0.000992000      1.237173000
Au      0.008259000      0.015831000      3.485680000
10  
0.862179920         0003.orco
Au      3.231274000      1.840249000     -1.010470000
Au      0.103630000     -1.879473000      0.325906000
Au     -3.412019000      1.477628000     -1.015893000
Au      0.131878000     -2.594485000      2.841337000
Au     -1.391220000     -0.164744000     -1.295869000
Au     -1.428680000      1.644133000      0.748463000
Au      0.120335000     -2.115788000     -2.346422000
Au      1.239622000      1.794337000      0.746257000
Au      1.399321000     -0.005832000     -1.293852000
Au      0.005860000      0.003976000      2.300544000
10  
0.895306180         0004.orco
Au     -0.000002000      2.647395000      2.265023000
Au     -0.000001000     -0.000002000     -4.250114000
Au      1.426978000      1.347265000      0.324033000
Au     -1.384658000     -0.000003000     -1.987186000
Au     -1.426974000     -1.347264000      0.324034000
Au      1.384663000     -0.000001000     -1.987191000
Au      1.426970000     -1.347262000      0.324032000
Au      0.000001000     -2.647393000      2.265015000
Au      0.000000000     -0.000003000      2.398320000
Au     -1.426976000      1.347267000      0.324033000
10  
1.038717304         0005.orco
Au     -1.009196000      1.658406000      3.330472000
Au      0.325149000     -1.883592000      0.000000000
Au     -1.009195000      1.658400000     -3.330467000
Au      2.841481000     -2.595592000     -0.000002000
Au     -1.295014000     -0.085585000     -1.397494000
Au      0.745711000      1.723387000     -1.336314000
Au     -2.347655000     -2.118901000     -0.000002000
Au      0.745712000      1.723384000      1.336313000
Au     -1.295019000     -0.085585000      1.397496000
Au      2.298026000      0.005678000     -0.000001000
10  
1.211232367         0006.orco
Au     -1.925619000      3.320841000      0.361492000
Au      1.615307000     -0.002510000     -1.017205000
Au     -1.936541000     -3.314310000      0.361419000
Au      3.838696000     -0.006970000      0.361397000
Au     -0.803891000     -1.395225000     -1.015347000
Au     -0.769329000     -1.336565000      1.708251000
Au     -0.000005000      0.000008000     -3.162514000
Au     -0.766724000      1.338271000      1.708218000
Au     -0.799482000      1.397857000     -1.015344000
Au      1.547587000     -0.001398000      1.709633000
10  
2.491258635         0007.orco
Au     -2.446512000     -2.639654000     -0.160102000
Au      1.386092000      0.231223000     -1.391187000
Au      3.555740000     -0.566279000     -0.157983000
Au     -1.259081000     -0.657887000     -1.393701000
Au     -1.153665000     -0.906596000      1.319085000
Au     -0.230473000      0.670389000     -3.490958000
Au     -0.620796000      1.793996000      1.481761000
Au     -0.693293000      2.024287000     -1.097432000
Au      1.479507000      0.014253000      1.327926000
Au     -0.017520000      0.036267000      3.562590000
10  
2.591208425         0008.orco
Au      0.000001000     -3.219414000      1.658932000
Au      3.540898000      0.000003000     -0.360535000
Au     -1.362981000      1.398064000      0.358973000
Au     -1.362981000     -1.398061000      0.358974000
Au      1.362984000     -1.398065000      0.358972000
Au     -3.540905000      0.000002000     -0.360539000
Au      1.362980000      1.398064000      0.358973000
Au      1.295901000      0.000000000     -2.016340000
Au     -1.295899000      0.000001000     -2.016336000
Au      0.000001000      3.219405000      1.658927000
10  
7.105854219         0009.orco
Au     -0.339550000      4.207975000     -0.680724000
Au     -1.174874000      1.796795000     -1.295024000
Au      0.680996000      2.161809000      0.667811000
Au     -2.379978000     -0.187502000     -2.518726000
Au     -1.277024000     -0.707470000     -0.211472000
Au     -1.447277000     -2.305872000      2.014018000
Au      0.752765000     -2.476352000      0.560354000
Au      3.003645000     -2.371466000     -0.750947000
Au      2.067236000     -0.060995000      0.168380000
Au      0.114062000     -0.056922000      2.046331000
10  
9.547708977         0010.orco
Au     -1.082935000     -1.390421000     -2.695453000
Au      1.246770000      1.360886000     -1.295967000
Au     -1.081318000      1.392458000     -2.695784000
Au     -1.145096000     -1.389360000     -0.032361000
Au     -2.643577000      0.000959000      1.738914000
Au     -1.144334000      1.389711000     -0.032810000
Au      1.245608000     -1.362450000     -1.296265000
Au      2.335226000     -0.001458000      0.754068000
Au      2.278545000     -0.000453000      3.419570000
Au     -0.008889000      0.000128000      2.136089000
10  
10.531349733        0011.orco
Au     -0.289003000      1.138980000     -1.784670000
Au      1.565093000      1.088028000     -3.638006000
Au     -0.402131000      1.422715000      1.009713000
Au     -2.580551000      1.742733000     -0.561863000
Au     -1.216262000     -2.179076000      2.020025000
Au     -1.629476000     -0.760336000     -0.196467000
Au      1.984639000     -0.309523000     -1.408796000
Au      0.679281000     -2.142480000      0.070036000
Au      1.881448000     -0.008754000      1.259932000
Au      0.006962000      0.007712000      3.230096000
-------------Gaussian code--------------------------- 
10    
0.000000000         mol_0001.out
Au  -2.277669000000  -2.652188000000   0.000322000000
Au  -0.312236000000  -1.355824000000  -1.431912000000
Au  -2.453526000000   0.000142000000  -0.000258000000
Au   1.997791000000  -0.000623000000  -1.393093000000
Au   4.262356000000   0.000023000000   0.000300000000
Au  -2.277268000000   2.652460000000  -0.000701000000
Au  -0.312450000000   1.355779000000   1.432101000000
Au  -0.312489000000  -1.355185000000   1.432322000000
Au   1.997404000000   0.000267000000   1.393035000000
Au  -0.311913000000   1.355149000000  -1.432116000000
10    
0.080947500         mol_0002.out
Au  -0.198569000000   1.537675000000  -1.712175000000
Au   1.496129000000  -0.620642000000   1.018693000000
Au   1.438123000000  -0.592585000000  -1.712408000000
Au  -0.004396000000  -0.003920000000   3.175815000000
Au  -1.224929000000  -0.944304000000  -1.714160000000
Au  -1.290229000000  -0.989065000000   1.017170000000
Au  -0.506630000000   3.818170000000  -0.362827000000
Au  -3.057040000000  -2.343974000000  -0.371952000000
Au  -0.216048000000   1.606512000000   1.019738000000
Au   3.563588000000  -1.467867000000  -0.357894000000
10    
0.106675000         mol_0003.out
Au  -2.281501000000  -0.360673000000   0.000000000000
Au   1.270395000000   0.284115000000   1.404793000000
Au  -0.470427000000  -1.820445000000   1.342982000000
Au   1.993223000000   2.472835000000   0.000000000000
Au  -0.470427000000  -1.820445000000  -1.342982000000
Au   1.270395000000   0.284115000000  -1.404793000000
Au  -3.228630000000   2.132730000000   0.000000000000
Au   1.270395000000  -1.491792000000  -3.337472000000
Au  -0.623818000000   1.811352000000   0.000000000000
Au   1.270395000000  -1.491792000000   3.337472000000
10    
0.410385000         mol_0004.out
Au  -1.432103000000   1.355486000000  -0.312258000000
Au   0.000000000000   0.000000000000  -2.453551000000
Au  -1.432103000000  -1.355486000000  -0.312258000000
Au   1.432103000000  -1.355486000000  -0.312258000000
Au   1.432103000000   1.355486000000  -0.312258000000
Au   0.000000000000   2.652308000000  -2.277491000000
Au  -1.393094000000   0.000000000000   1.997609000000
Au   1.393094000000   0.000000000000   1.997609000000
Au   0.000000000000   0.000000000000   4.262347000000
Au   0.000000000000  -2.652308000000  -2.277491000000
10    
0.411012500         mol_0005.out
Au   1.432113000000   1.355499000000  -0.312278000000
Au   0.000000000000   0.000000000000  -2.453464000000
Au   0.000000000000  -2.652338000000  -2.277470000000
Au  -1.432113000000  -1.355499000000  -0.312278000000
Au  -1.393063000000   0.000000000000   1.997582000000
Au   1.432113000000  -1.355499000000  -0.312278000000
Au   1.393063000000   0.000000000000   1.997582000000
Au  -1.432113000000   1.355499000000  -0.312278000000
Au   0.000000000000   0.000000000000   4.262351000000
Au   0.000000000000   2.652338000000  -2.277470000000
10    
0.411012500         mol_0006.out
Au   0.000000000000   1.393077000000   1.997577000000
Au   1.355502000000  -1.432130000000  -0.312278000000
Au   0.000000000000   0.000000000000   4.262334000000
Au  -1.355502000000   1.432130000000  -0.312278000000
Au   0.000000000000   0.000000000000  -2.453446000000
Au   1.355502000000   1.432130000000  -0.312278000000
Au   2.652332000000   0.000000000000  -2.277464000000
Au  -1.355502000000  -1.432130000000  -0.312278000000
Au   0.000000000000  -1.393077000000   1.997577000000
Au  -2.652332000000   0.000000000000  -2.277464000000
10    
0.605537500         mol_0007.out
Au  -1.142989000000  -2.686582000000   1.381098000000
Au   1.224943000000  -1.317578000000   1.372248000000
Au   0.035859000000   2.137292000000   0.000000000000
Au  -1.142989000000  -0.008310000000  -1.404005000000
Au  -1.142989000000  -0.008310000000   1.404005000000
Au   2.347618000000   3.384638000000   0.000000000000
Au   2.363268000000   0.711248000000   0.000000000000
Au  -1.142989000000  -2.686582000000  -1.381098000000
Au   1.224943000000  -1.317578000000  -1.372248000000
Au  -2.624675000000   1.791761000000   0.000000000000
10    
0.651972500         mol_0008.out
Au   1.142222000000  -0.008544000000   1.403927000000
Au   1.142222000000  -0.008544000000  -1.403927000000
Au   2.625562000000   1.790229000000   0.000000000000
Au  -0.034592000000   2.138486000000   0.000000000000
Au  -2.361993000000   0.712122000000   0.000000000000
Au  -1.225679000000  -1.317760000000  -1.372283000000
Au  -1.225679000000  -1.317760000000   1.372283000000
Au   1.142222000000  -2.686842000000   1.381130000000
Au   1.142222000000  -2.686842000000  -1.381130000000
Au  -2.346507000000   3.385456000000   0.000000000000
10    
0.655110000         mol_0009.out
Au  -1.225885000000  -1.317773000000   1.372289000000
Au   1.142012000000  -2.686900000000   1.381090000000
Au  -2.361718000000   0.712384000000   0.000000000000
Au   1.142012000000  -2.686900000000  -1.381090000000
Au   1.142012000000  -0.008583000000  -1.403959000000
Au   1.142012000000  -0.008583000000   1.403959000000
Au  -0.034252000000   2.138674000000   0.000000000000
Au  -1.225885000000  -1.317773000000  -1.372289000000
Au  -2.346136000000   3.385700000000   0.000000000000
Au   2.625828000000   1.789754000000   0.000000000000
10    
0.665777500         mol_0010.out
Au  -1.141723000000  -2.686990000000   1.381069000000
Au   1.226160000000  -1.317807000000   1.372307000000
Au   1.226160000000  -1.317807000000  -1.372307000000
Au   2.361299000000   0.712719000000   0.000000000000
Au  -1.141723000000  -0.008657000000  -1.403962000000
Au  -1.141723000000  -2.686990000000  -1.381069000000
Au  -2.626177000000   1.789138000000   0.000000000000
Au  -1.141723000000  -0.008657000000   1.403962000000
Au   2.345670000000   3.386019000000   0.000000000000
Au   0.033782000000   2.139031000000   0.000000000000
10    
0.677700000         mol_0011.out
Au  -2.344522000000   3.386742000000   0.000000000000
Au  -1.226774000000  -1.317819000000   1.372367000000
Au   1.141057000000  -0.008795000000   1.404001000000
Au   1.141057000000  -2.687179000000   1.380962000000
Au  -1.226774000000  -1.317819000000  -1.372367000000
Au  -0.032713000000   2.139695000000   0.000000000000
Au   1.141057000000  -0.008795000000  -1.404001000000
Au   2.627011000000   1.787673000000   0.000000000000
Au   1.141057000000  -2.687179000000  -1.380962000000
Au  -2.360455000000   0.713474000000   0.000000000000
10    
0.680837500         mol_0012.out
Au  -1.141334000000  -0.008739000000   1.403967000000
Au  -2.626669000000   1.788302000000   0.000000000000
Au   1.226521000000  -1.317840000000   1.372316000000
Au  -1.141334000000  -2.687089000000   1.381050000000
Au   1.226521000000  -1.317840000000  -1.372316000000
Au  -1.141334000000  -2.687089000000  -1.381050000000
Au  -1.141334000000  -0.008739000000  -1.403967000000
Au   0.033151000000   2.139435000000   0.000000000000
Au   2.360804000000   0.713162000000   0.000000000000
Au   2.345008000000   3.386439000000   0.000000000000
10    
0.766805000         mol_0013.out
Au   0.001696000000   1.550698000000  -1.713105000000
Au   1.404204000000  -0.810632000000   1.018295000000
Au   1.342096000000  -0.776818000000  -1.713105000000
Au   0.000000000000   0.000000000000   3.176085000000
Au  -1.343792000000  -0.773881000000  -1.713105000000
Au  -1.404130000000  -0.810760000000   1.018295000000
Au   0.000000000000   3.851736000000  -0.363885000000
Au  -3.335701000000  -1.925868000000  -0.363885000000
Au  -0.000074000000   1.621393000000   1.018295000000
Au   3.335701000000  -1.925868000000  -0.363885000000
10    
0.770570000         mol_0014.out
Au   0.000000000000   1.550762000000  -1.713204000000
Au   1.404981000000  -0.811166000000   1.018541000000
Au   1.342999000000  -0.775381000000  -1.713204000000
Au   0.000000000000   0.000000000000   3.176014000000
Au  -1.342999000000  -0.775381000000  -1.713204000000
Au  -1.404981000000  -0.811166000000   1.018541000000
Au   0.000000000000   3.852676000000  -0.364009000000
Au  -3.336515000000  -1.926338000000  -0.364009000000
Au   0.000000000000   1.622332000000   1.018541000000
Au   3.336515000000  -1.926338000000  -0.364009000000
10    
0.886657500         mol_0015.out
Au  -3.558900000000  -0.000505000000   0.368920000000
Au  -1.364617000000  -1.401878000000  -0.352941000000
Au   1.298566000000  -0.000025000000   2.051994000000
Au  -1.297737000000  -0.000096000000   2.052445000000
Au   3.559039000000   0.000482000000   0.367534000000
Au  -1.365262000000   1.401712000000  -0.353025000000
Au   1.364453000000   1.401887000000  -0.353347000000
Au   0.000041000000  -3.185079000000  -1.714103000000
Au   1.365097000000  -1.401690000000  -0.353584000000
Au  -0.000678000000   3.185194000000  -1.713893000000
10    
1.297670000         mol_0016.out
Au  -3.185140000000   0.000000000000  -1.713996000000
Au   0.000000000000   1.298151000000   2.052218000000
Au   0.000000000000  -1.298151000000   2.052218000000
Au  -1.401792000000  -1.364858000000  -0.353227000000
Au   0.000000000000   3.558968000000   0.368233000000
Au   1.401792000000  -1.364858000000  -0.353227000000
Au   1.401792000000   1.364858000000  -0.353227000000
Au  -1.401792000000   1.364858000000  -0.353227000000
Au   0.000000000000  -3.558968000000   0.368233000000
Au   3.185140000000   0.000000000000  -1.713996000000
10    
3.623812500         mol_0017.out
Au  -0.000133000000  -4.044760000000   0.000000000000
Au   2.361091000000  -2.681878000000   0.000000000000
Au  -0.000003000000  -1.356045000000   0.000000000000
Au  -2.361181000000  -2.682114000000   0.000000000000
Au   2.375255000000   0.000175000000   0.000000000000
Au   2.361212000000   2.682121000000   0.000000000000
Au   0.000000000000   1.356013000000   0.000000000000
Au  -2.375273000000  -0.000133000000   0.000000000000
Au  -2.361104000000   2.681914000000   0.000000000000
Au   0.000135000000   4.044708000000   0.000000000000
10    
3.635107500         mol_0018.out
Au  -1.356701000000  -0.000001000000   0.000727000000
Au   1.356700000000   0.000004000000  -0.000451000000
Au  -0.000001000000  -2.375274000000   0.000020000000
Au  -0.000001000000   2.375276000000  -0.000036000000
Au  -2.682473000000  -2.361949000000  -0.000022000000
Au  -2.682473000000   2.361948000000   0.000005000000
Au  -4.045400000000   0.000000000000  -0.000292000000
Au   2.682469000000  -2.361944000000  -0.000071000000
Au   2.682473000000   2.361943000000  -0.000044000000
Au   4.045407000000  -0.000004000000   0.000165000000
10    
3.868537500         mol_0019.out
Au   0.776682000000  -0.647616000000  -2.020257000000
Au  -0.216072000000  -1.415974000000   0.547985000000
Au   1.392659000000  -1.906182000000   2.569742000000
Au   2.472326000000  -0.627633000000   0.547183000000
Au   3.087532000000   0.852954000000  -1.555199000000
Au   0.755352000000   1.464198000000  -0.225896000000
Au  -0.585969000000   2.965742000000   1.444515000000
Au  -1.801271000000  -1.051254000000  -1.578726000000
Au  -1.883686000000   0.932087000000   0.384578000000
Au  -3.997554000000  -0.566322000000  -0.113925000000
10    
4.441445000         mol_0020.out
Au   0.000000000000   0.000000000000   1.355480000000
Au   0.000000000000   2.360610000000   2.682075000000
Au   0.000000000000  -2.376378000000   0.000000000000
Au   0.000000000000   0.000000000000  -1.355480000000
Au   0.000000000000   0.000000000000  -4.044978000000
Au   0.000000000000   2.376378000000   0.000000000000
Au   0.000000000000  -2.360610000000   2.682075000000
Au   0.000000000000  -2.360610000000  -2.682075000000
Au   0.000000000000   0.000000000000   4.044978000000
Au   0.000000000000   2.360610000000  -2.682075000000
10    
6.411167500         mol_0021.out
Au  -3.306659000000  -2.609925000000   0.000158000000
Au  -3.303762000000   0.000000000000  -0.000474000000
Au  -0.932438000000   1.435071000000   0.000329000000
Au  -3.306659000000   2.609926000000   0.000144000000
Au   1.430200000000   2.677719000000  -0.000320000000
Au   3.775262000000   1.338781000000   0.000156000000
Au  -0.932439000000  -1.435071000000   0.000284000000
Au   1.430198000000  -2.677718000000  -0.000325000000
Au   1.371034000000   0.000000000000  -0.000103000000
Au   3.775262000000  -1.338783000000   0.000152000000
10    
6.824062500         mol_0022.out
Au   0.000000000000   0.000000000000  -1.371091000000
Au   0.000000000000   1.434786000000   0.932372000000
Au   0.000000000000   2.610176000000   3.306399000000
Au   0.000000000000   1.338906000000  -3.775131000000
Au   0.000000000000   0.000000000000   3.303887000000
Au   0.000000000000   2.677481000000  -1.430038000000
Au   0.000000000000  -2.677481000000  -1.430038000000
Au   0.000000000000  -1.338906000000  -3.775131000000
Au   0.000000000000  -1.434786000000   0.932372000000
Au   0.000000000000  -2.610176000000   3.306399000000
10    
6.824062500         mol_0023.out
Au   0.000000000000   1.434812000000   0.932383000000
Au   0.000000000000   0.000000000000  -1.371055000000
Au   0.000000000000   2.610180000000   3.306419000000
Au   0.000000000000  -1.434812000000   0.932383000000
Au   0.000000000000   2.677473000000  -1.430064000000
Au   0.000000000000  -1.338900000000  -3.775147000000
Au   0.000000000000  -2.677473000000  -1.430064000000
Au   0.000000000000   1.338900000000  -3.775147000000
Au   0.000000000000  -2.610180000000   3.306419000000
Au   0.000000000000   0.000000000000   3.303871000000
\end{verbatim}
\typeout{get arXiv to do 4 passes: Label(s) may have changed. Rerun}
\end{document}